\documentstyle[12pt]{article}

\begin{document}

%%%%%%%%%%%%%%%%%%%%%%%%%%%%%%%%%%%%%%%%%%%%%%%%%%%%%%%%%%%%%%%%%%%%
%%%%%%%%%%%%%%%%%%%%%%%%%%%%%%  Defs. %%%%%%%%%%%%%%%%%%%%%%%%%%%%%%
%%%%%%%%%%%%%%%%%%%%%%%%%%%%%%%%%%%%%%%%%%%%%%%%%%%%%%%%%%%%%%%%%%%%

\def\a{\alpha}
\def\b{\beta}
\def\c{\varepsilon}
\def\d{\delta}
\def\e{\epsilon}
\def\f{\phi}
\def\g{\gamma}
\def\h{\theta}
\def\k{\kappa}
\def\l{\lambda}
\def\m{\mu}
\def\n{\nu}
\def\p{\psi}
\def\q{\partial}
\def\r{\rho}
\def\s{\sigma}
\def\t{\tau}
\def\u{\upsilon}
\def\v{\varphi}
\def\w{\omega}
\def\x{\xi}
\def\y{\eta}
\def\z{\zeta}
\def\D{\Delta}
\def\G{\Gamma}
\def\L{\Lambda}
\def\F{\Phi}
\def\P{\Psi}
\def\S{\Sigma}

\def\o{\over}

\def\IJMP{Int.~J.~Mod.~Phys. }
\def\MPL{Mod.~Phys.~Lett. }
\def\NP{Nucl.~Phys. }
\def\PL{Phys.~Lett. }
\def\PR{Phys.~Rev. }
\def\PRL{Phys.~Rev.~Lett. }
\def\PTP{Prog.~Theor.~Phys. }
\def\ZP{Z.~Phys. }

\def\beq{\begin{equation}}
\def\eeq{\end{equation}}

%%%%%%%%%%%%%%%%%%%%%%%%%%%%%%%%%%%%%%%%%%%%%%%%%%%%%%%%%%%%%%%%%%%%

\title{
\begin{flushright}
\large UT-770
\end{flushright}
       $R$-invariant Natural Unification}

\author{Izawa K.-I. and T. Yanagida \\
        \\ {\sl Department of Physics, University of Tokyo, Tokyo
        113, Japan}}

\date{March, 1997}

\maketitle\thispagestyle{empty}
\setlength{\baselineskip}{3.6ex}

\begin{abstract}
We construct $R$-invariant unification models
where a pair of massless Higgs doublets is naturally obtained.
The masslessness of the Higgs doublets is guaranteed by the unbroken
$R$ symmetry.
Mass generation for the Higgs doublets is considered
from various viewpoints.
\end{abstract}

\newpage

\section{Introduction}

Nonvanishing superpotential gives a negative cosmological constant in
supergravity. $R$ symmetry is a unique symmetry that forbids a constant
term in superpotential and thus it may play a fundamental role for
understanding a vanishing cosmological constant in supergravity.
The $R$ symmetry has been widely considered in phenomenology of supersymmetric
(SUSY) gauge theories or supergravity, since it (or its discrete subgroup)
can avoid too rapid proton decays
\cite{Dim, Wei}
and provide a candidate for cold dark
matter in our universe
\cite{Who}.
In a recent article
\cite{Iza},
it has been pointed out that
the spontaneous breakdown of the $R$ symmetry U$(1)_R$ to its discrete
subgroup Z$_{2nR}$
\cite{Kum}
produces a flat potential for a new inflation model.

Motivated by the above theoretical and phenomenological arguments,
we construct, in this paper, $R$-invariant unification models.

In the next section, we show that the minimal SUSY grand unified theory
(GUT) is easily extended to an $R$-invariant one.
However, we stress that this model has a serious doublet-triplet
splitting problem as all the SUSY-GUT's do.
In section three, we construct an $R$-invariant extension of recently
proposed natural unification theories
\cite{Yan, Hot},
where the doublet-triplet splitting
problem is solved. Namely, masslessness of the Higgs doublets is
guaranteed by the $R$ symmetry, while the Higgs triplets have $R$-invariant
masses at the unification scale. In this model, however, a pair of Higgs
doublets is completely massless as long as the $R$ symmetry is unbroken.
In section four, we discuss how to generate a mass for the Higgs doublets
at the electroweak scale by modifying the above model.
The final section is devoted to a discussion on low-energy predictions
of the $R$-invariant natural unification model. A possible connection
to the superstring theory is also briefly noted.

\section{A SUSY-GUT with a U$(1)_R$ symmetry}

In this section, we consider the minimal GUT based on a gauge group SU(5)
\cite{Geo}.
The Higgs sector in the SUSY SU(5) GUT consists of a pair of Higgs
chiral multiplets $H_i$ and ${\bar H}^i$
\cite{Fay}
transforming as ${\bf 5}$ and ${\bf 5}^*$
of SU$(5)_{GUT}$ and a ${\bf 24}$ Higgs
chiral multiplet $\S^i_j$ ($i, j = 1, \cdots, 5$).
The vacuum expectation value of the adjoint Higgs $\S$ is supposed to
break the SU$(5)_{GUT}$ group down to the standard-model gauge group
${\rm SU}(3)_C \times {\rm SU}(2)_L \times {\rm U}(1)_Y$:
\begin{equation}
  \label{VEV5}
  \begin{array}{c}
    \displaystyle \langle \S \rangle =
    \left(
      \begin{array}{ccccc}
        2 & & & & \\
        & 2 & & & \\
        & & 2 & & \\
        & & & -3 & \\
        & & & & -3
      \end{array}
    \right)
    V, \\
  \end{array}
\end{equation}
where $V$ is at the GUT scale of order $10^{16}$ GeV.

An important point is that $R$ charge of $\S$ must be vanishing,
otherwise the $R$ symmetry is spontaneously broken at the GUT scale
to produce a negative cosmological constant of order the GUT scale
in supergravity. Provided that the negative vacuum energy is canceled out
by condensation energy of SUSY-breaking in the hidden sector,
we obtain too large SUSY-breaking scale of order $10^{14}$ GeV
in the visible sector
\cite{Iza}.
Therefore we assume that the $\S$ multiplet
transforms trivially under the U$(1)_R$ symmetry.

In order to construct a superpotential with $R$ charge two,
we introduce another adjoint Higgs $\S'^i_j$ whose $R$ charge is two.
Then the renormalizable superpotential for the adjoint Higgs
is given by
\beq
W_{\S} = m{\rm Tr}(\S' \S) + \l{\rm Tr}(\S' \S^2).
\eeq
We have a desired SUSY vacuum:
\beq
\langle \S' \rangle = 0
\label{VEV}
\eeq
with Eq.(\ref{VEV5}) and $V = m/\l$.
In this vacuum, the GUT gauge group is broken down to the standard-model one,
while the superpotential is kept vanishing: $\langle W \rangle = 0$.
This results from the fact that the $R$ symmetry U$(1)_R$ is unbroken under
Eqs.(\ref{VEV}) and (\ref{VEV5}).

Let us turn to the mass term for the Higgs $H$ and ${\bar H}$.
We assume that the product $H{\bar H}$ has $R$ charge two to
make the color triplets in $H$ and ${\bar H}$ heavy.
Then their superpotential is given by
\beq
W_H = m'H{\bar H} + \l'H\S{\bar H}.
\eeq
In the broken phase Eq.(\ref{VEV5}), the color-triplet and weak-doublet
components of $H$ and ${\bar H}$ have different masses as
\beq
m_{H_C} = m' + {2\l' \o \l}m, \quad m_{H_f} = m' - {3\l' \o \l}m,
\eeq
respectively.
As pointed out in Ref.\cite{Dim, Sak}, we need an extreme fine-tuning
even in the SUSY extension of GUT's to obtain a pair of Higgs
doublets with $m_{H_f}$ of order $10^2$ GeV.
To avoid this unnatural tuning, we proceed to consider
$R$ invariance in recently proposed natural unification scheme
\cite{Yan}.

\section{An $R$-invariant natural unification model}

The natural unification models
\cite{Yan, Hot}
are based on a product of two
distinct gauge groups, $G_{GUT}$ and $G_H$.
The group $G_{GUT}$ is the usual GUT gauge group and its
coupling constant
is in a perturbative regime, while the group $G_H$ is a hypercolor
gauge group whose coupling is strong at the unification scale.
The color gauge group SU$(3)_C$ (or ${\rm SU}(3)_C \times {\rm U}(1)_Y$)
at low energies is a linear combination of SU(3) (or SU(3) $\times$ U(1))
subgroups of $G_{GUT}$ and $G_H$, and the weak gauge group SU$(2)_L$
is a subgroup of $G_{GUT}$.
\footnote{The converse choice is also possible.}
The strong coupling for the hypercolor group
$G_H$ is necessary to achieve an approximate unification of
three gauge coupling constants of
${\rm SU}(3)_C \times {\rm SU}(2)_L \times {\rm U}(1)_Y$.

In this paper, we consider an ${\rm SU}(5)_{GUT} \times {\rm U}(3)_H$ model.
The quarks and leptons obey the usual transform law
\cite{Geo}
under the GUT group SU$(5)_{GUT}$, while they are all singlets of the
hypercolor group U$(3)_H$.
We introduce a pair of Higgs $H_i$ and ${\bar H}^i$ ($i = 1, \cdots, 5$)
which transform as ${\bf 5}$ and ${\bf 5}^*$ under the SU$(5)_{GUT}$
and as singlets under the U$(3)_H$.
So far all matter multiplets are the same as
in the minimal SUSY SU$(5)_{GUT}$.
This guarantees the electric charge quantization for the ordinary sector
and the $m_b = m_\t$ unification.

We introduce six pairs of hyperquarks $Q_\a^I$ and ${\bar Q}_I^\a$
($\a = 1, \cdots, 3; I = 1, \cdots, 6$) which transform as ${\bf 3}$
and ${\bf 3}^*$ under the hypercolor SU$(3)_H$
and have U$(1)_H$ charges $1$ and $-1$, respectively.
The first five pairs $Q_\a^i$ and ${\bar Q}_i^\a$ belong to ${\bf 5}^*$
and ${\bf 5}$ of SU$(5)_{GUT}$, respectively, and the last pair $Q_\a^6$
and ${\bar Q}_6^\a$ are singlets of SU$(5)_{GUT}$.

Since $Q_\a^i$ and ${\bar Q}_i^\a$ are supposed to have vacuum expectation
values of order the unification scale, they must be trivial
representations of the $R$ symmetry U$(1)_R$ as explained
in the previous section.
To cause a desired breaking of the total gauge group
${\rm SU}(5)_{GUT} \times {\rm U}(3)_H$, we are led to introduce chiral multiple
with $R$ charges two which couple to $Q_\a^I$ and ${\bar Q}_I^\a$.
We take an adjoint representation $X^\a_\b$
\cite{His}
of U$(3)_H$ in this paper.
The renormalizable superpotential for $Q_\a^I$, ${\bar Q}_I^\a$,
$X^\a_\b$, $H_i$, and ${\bar H}^i$ is given by
\beq
W = \l Q_\a^i {\bar Q}_i^\b X^\a_\b + \l' Q_\a^6 {\bar Q}_6^\b X^\a_\b
    + h Q_\a^i {\bar Q}_6^\a H_i + h' Q_\a^6 {\bar Q}_i^\a {\bar H}^i,
\label{POT}
\eeq
where we have assumed the $R$ charges of $H_i$ and ${\bar H}^i$ to be two
and those of $Q_\a^6$ and ${\bar Q}^\a_6$ vanishing.
In addition to the U$(1)_R$, this superpotential possesses another
axial symmetry U$(1)_\chi$ given by
\beq
Q_\a^I \rightarrow e^{i\x}Q_\a^I, \quad
{\bar Q}^\a_I \rightarrow e^{i\x}{\bar Q}^\a_I, \quad
X^\a_\b \rightarrow e^{-2i\x}X^\a_\b, \quad
H_i \rightarrow e^{-2i\x}H_i, \quad
{\bar H}^i \rightarrow e^{-2i\x}{\bar H}^i.
\eeq
We can also impose this axial symmetry to avoid nonrenormalizable terms
in the superpotential.
Note that the global symmetries U$(1)_R$ and U$(1)_\chi$ have no SU$(3)_H$
anomaly.

The regular terms in the effective superpotential
allowed by the symmetries
\cite{Sei}
of the model is exactly
the tree-level ones in Eq.(\ref{POT}).
\footnote{An invariant term with $R$ charge two given by
$\{\det(Q^I_\a X^\a_\b {\bar Q}^\b_J)\}^{1/6}$ is singular at the origin
$Q^I_\a = {\bar Q}^\a_I = X^\a_\b = 0$.}
Thus we consider the tree-level
$D$ and $F$ term flatness conditions to obtain our quantum vacua.

The $F$-flatness conditions are given by
\beq
\begin{array}{l}
\displaystyle
\l \{Q_\a^i{\bar Q}_i^\b - {1 \o 3}\d_\a^\b{\rm Tr}(Q^i{\bar Q}_i)\}
+ \l' \{ Q_\a^6 {\bar Q}_6^\b - {1 \o 3}\d_\a^\b{\rm Tr}(Q^6{\bar Q}_6)\} = 0,
\\
\noalign{\vskip 2ex}
\displaystyle
Q_\a^i{\bar Q}^\a_6 = 0, \quad Q_\a^6{\bar Q}_i^\a = 0, \quad
\l {\bar Q}_i^\a X_\b^\a + h {\bar Q}^\a_6 H_i = 0,
\\
\noalign{\vskip 2ex}
\displaystyle
\l Q^i_\a X_\b^\a + h' Q_\b^6 {\bar H}^i = 0, \quad
\l' {\bar Q}_6^\b X_\b^\a + h' {\bar Q}^\a_i {\bar H}^i = 0, \quad
\l' Q^6_\a X_\b^\a + h Q_\b^i H_i = 0.
\end{array}
\eeq
Together with the $D$-flatness conditions for U$(3)_H$,
we obtain desired vacua as follows:
\beq
\langle X^\a_\b \rangle = \langle H_i \rangle = \langle {\bar H}^i \rangle = 0,
\quad
\langle Q^6_\a \rangle = \langle {\bar Q}^\a_6 \rangle = 0, \quad
\langle Q^i_\a \rangle = v\d^i_\a, \quad
\langle {\bar Q}^\a_i \rangle = v\d^\a_i.
\label{VAC}
\eeq
For $v \neq 0$, the total gauge group ${\rm SU}(5)_{GUT} \times {\rm U}(3)_H$
is broken down to ${\rm SU}(3)_C \times {\rm SU}(2)_L \times {\rm U}(1)_Y$.

As noted in the beginning of this section, the color SU$(3)_C$
and the U$(1)_Y$ are, respectively, a linear combination of an SU(3) subgroup
of the SU$(5)_{GUT}$ and the hypercolor SU$(3)_H$ and that of a U(1)
subgroup of the SU$(5)_{GUT}$ and the strong U$(1)_H$.
Thus the gauge coupling constants $\a_C$, $\a_2$, and $\a_1$
for ${\rm SU}(3)_C \times {\rm SU}(2)_L \times {\rm U}(1)_Y$
are given by
\cite{Yan}
\beq
\a_C \simeq {\a_{GUT} \o 1+\a_{GUT}/\a_{3H}}, \quad
\a_2 = \a_{GUT}, \quad
\a_1 \simeq {\a_{GUT} \o 1+{1 \o 15}\a_{GUT}/\a_{1H}},
\label{COU}
\eeq
where $\a_{3H}$ and $\a_{1H}$ denote gauge coupling constants
\footnote{See Ref.\cite{Yan} for the normalization of $\a_{1H}$.}
for the hypercolor SU$(3)_H$ and U$(1)_H$, respectively.
We see from Eq.(\ref{COU}) that the unification of three gauge
coupling constants $\a_C, \a_2, \a_1$ is achieved in the strong coupling
limit of the hypercolor gauge interaction:
$\a_{3H}$ and $\a_{1H} \rightarrow \infty$.
\footnote{Precisely speaking, in the strong coupling region,
$\a_{3H}, \a_{1H} \gg 1$, the threshold corrections may yield substantial
effects on the relations in Eq.(\ref{COU}).
However, it has been pointed out
\cite{Mor}
that $\a_{3H}, \a_{1H} \simeq {\cal O}(1)$ is sufficient to get
the observed values of $\a_C$, $\a_2$, and $\a_1$.}

In the vacuum Eq.(\ref{VAC}), the color triplets $H_a$ and ${\bar H}^a$
($a = 1, \cdots, 3$) gain masses of order $v$ together with the sixth
hyperquarks ${\bar Q}^\a_6$ and $Q^6_\a$ (see Eq.(\ref{POT})).
On the other hand, the weak doublets $H_i$ and ${\bar H}^i$ ($i = 4, 5$)
remain massless since there are no partners for them.
The masslessness for these doublets is guaranteed by the unbroken $R$ symmetry.

Notice that the scale $v$ is undetermined so far. This implies the presence of
a flat direction in the present vacuum. The imaginary part of the scalar compone
of the corresponding massless chiral multiplet is a Nambu-Goldstone
mode related to the breaking of the axial U$(1)_\chi$ symmetry.

Thus we should break
the U$(1)_\chi$ explicitly to fix the unification scale $v$
and eliminate the flat direction.
We introduce a singlet chiral multiplet $\f$ whose U$(1)_R$ and U$(1)_\chi$
charges are $2$ and $-2$ to have a soft breaking of the axial U$(1)_\chi$.
Then we have a superpotential for $\f$:
\beq
W_\f = k Q^i_\a {\bar Q}^\a_i \f + k' Q^6_\a {\bar Q}^\a_6 \f + M_\f^2 \f.
\label{FPO}
\eeq
The $M_\f^2$ term is a U$(1)_\chi$ breaking one. With this superpotential,
we obtain $3kv^2 = M_\f^2$.

The U$(1)_\chi$-breaking mass $M_\f$ may stem from a condensation of some
other hypercolor quarks. For example, we consider an SU$(2)_{H'}$ strong
gauge theory with four hyperquarks ${Q'}_\a^i$
where $\a = 1, 2$ and $i = 1, \cdots, 4$.
Provided that the hyperquarks ${Q'}_\a^i$ have vanishing $R$ charges and
their U$(1)_\chi$ charges are one, the singlet $\f$ can couple to ${Q'}_\a^i$ as
\beq
W'_\f = k''_{ij} \e^{\a \b} {Q'}_\a^i {Q'}_\b^j \f.
\eeq
Nonperturbative effects of the strong SU$(2)_{H'}$ cause
the $\e^{\a \b} {Q'}_\a^i {Q'}_\b^j$ condensation
\cite{Kap},
which eventually induces the $M_\f^2$ term
in Eq.(\ref{FPO}).

Since the U$(1)_\chi$ has the strong SU$(2)_{H'}$ anomaly,
there arises no massless Nambu-Goldstone multiplet
associated with the spontaneous breaking of U$(1)_\chi$.
Note that the $R$ symmetry U$(1)_R$ has no strong SU$(2)_{H'}$ anomaly
and hence it is still an exact symmetry.

We finally stress that there is no massless multiplet in the present vacuum
except for the pair of Higgs doublets $H_i$ and ${\bar H}^i$  ($i = 4, 5$) and
three families of quark-lepton chiral multiplets.
Our low-energy spectrum is nothing other than that of the SUSY standard model.

\section{Mass generation for the Higgs doublets}

The Higgs doublets are kept massless by the U$(1)_R$ symmetry.
All the quark-lepton chiral multiplets have vanishing U$(1)_R$ charges
so that they have Yukawa couplings
to the Higgs doublets $H_i$ and ${\bar H}^i$.
It is remarkable that the $R$ symmetry has a QCD anomaly and thus plays
a role of the Peccei-Quinn symmetry
\cite{Pec}
to suppress the strong CP violation.

The breaking scale $v_{PQ}$ of the Peccei-Quinn symmetry is subject to
astrophysical and cosmological constraints
\cite{Kim}
as
\beq
10^{10} {\rm GeV} \leq v_{PQ} \leq 10^{12} {\rm GeV}.
\eeq
Thus we assume that the breaking scale $v_R$ of the U$(1)_R$ symmetry
satisfies the same constraints:
\beq
10^{10} {\rm GeV} \leq v_R \leq 10^{12} {\rm GeV}.
\label{INT}
\eeq
In general, the Higgs doublets receive masses
from the U$(1)_R$ breaking sector. The value of their masses depends on the $R$
charge of the Higgs field $\y$ which breaks the $R$ symmetry.

When the $\y$ has $R$ charge $-1$, a nonrenormalizable term
\beq
W_\h = {g \o M} H_i {\bar H}^i \y^2
\eeq
induces a Higgs mass
\beq
\m \simeq g {{\langle \y \rangle}^2 \o M} \simeq g {v_R^2 \o M},
\label{HMM}
\eeq
where $M$ denotes
the gravitational scale $M \simeq 2.4 \times 10^{18} {\rm GeV}$.
Eqs.(\ref{INT}) and (\ref{HMM}) lead to a mass of the electroweak scale
for the Higgs doublets with $g$ of order one.
This may be an encouraging result in the present model,
though it seems accidental that the Higgs mass $\m$
comes out to be of the same order as the SUSY-breaking scale.

In the rest of this section, we consider another model
so modified that the Giudice-Masiero term
\cite{Giu}
induces the Higgs-doublet mass
of the electroweak scale through SUSY breaking in the hidden sector.
\footnote{One may keep only a discrete subgroup of U$(1)_R$
in the previous model to allow the Giudice-Masiero term.}
The difference between the previous and modified models
amounts to the global U(1) charge assignments.
The charges in the two models are given in Table \ref{tbl: charges}.
In the modified model,
the renormalizable superpotential is given by
\beq
W = \l Q_\a^i {\bar Q}^\b_i X_\b^\a + k Q_\a^i {\bar Q}^\a_i \f
    + h Q_\a^i {\bar Q}^\a_6 H_i + h' Q_\a^6 {\bar Q}^\a_i {\bar H}^i
    + M_\f^2 \f,
\label{POTMOD}
\eeq
where the $M_\f^2$ term is the soft breaking one of the U$(1)_\chi$ symmetry.
We get the desired vacuum as in the previous model.
\footnote{The superpotential Eq.(\ref{POTMOD}) possesses another global
U(1). Two linear combinations of the three global U(1)'s are the same as
the U$(1)_R$ and U$(1)_\chi$ in the previous model. One must introduce
nonrenormalizable superpotentials to eliminate the extra U(1), which
however does not affect our conclusion.}

The Higgs doublets $H_i$ and ${\bar H}^i$ ($i = 4, 5$)
never acquire masses as long as
$\langle X \rangle = \langle \f \rangle = \langle Q_\a^6 \rangle
= \langle {\bar Q}_6^\a \rangle = 0$.
In contrast to the previous model,
however, the Higgs multiplets $H_i$ and ${\bar H}^i$ have vanishing charges
for U$(1)_R$ and U$(1)_\chi$, which allows their coupling to the Polonyi
field $Z$ in the K{\" a}hler potential as
\beq
K = {g' \o M} Z^* H_i {\bar H}^i + h.c.
\label{KAH}
\eeq
Since the Polonyi field $Z$ is supposed to have a nonvanishing $F$ term
$\langle F_Z \rangle \simeq (10^{11} {\rm GeV})^2$ to break the SUSY,
the K{\" a}hler potential (\ref{KAH}) gives rise to the Higgs mass
\cite{Giu}
\beq
\m \simeq g' {\langle F_Z \rangle \o M}.
\eeq
With $g'$ of order one, the mass $\m$ turns out to be of the SUSY-breaking
scale or the electroweak scale as desired.

\section{Conclusion}

We have constructed $R$-invariant unification models
where a pair of massless Higgs doublets is naturally obtained.
\footnote{We have used U$(1)_R$ in this paper, though a discrete
$R$ symmetry $Z_{nR}$ with large $n$ is sufficient for our purpose.}
The masslessness of the Higgs doublets is guaranteed by the unbroken
$R$ symmetry.
These natural unification models are based on the gauge group
${\rm SU}(5)_{GUT} \times {\rm U}(3)_H$. All of the quark-lepton multiplets
and the Higgs $H_i$ and ${\bar H}^i$ ($i = 1, \cdots, 5$) are singlets
of the U$(3)_H$ and they belong to the standard representations of the
SU$(5)_{GUT}$ as in the minimal SUSY-GUT.
Therefore some predictions on the quark-lepton sector are intact
in our models with a non-simple gauge group.
The $m_b = m_\t$ unification is an example.
Another example is the unification of soft SUSY-breaking masses for
scalar quarks and leptons. Namely, scalar components of the chiral multiplets
which belong to to the same multiplets of the SU$(5)_{GUT}$ have the same
SUSY-breaking masses at the unification scale.

On the contrary, the gauge sector in the natural unification models
is different from that in the usual SUSY-GUT's.
It is a crucial difference that the color SU$(3)_C$ is a diagonal
subgroup of an SU(3) subgroup of the SU$(5)_{GUT}$
and the hypercolor SU$(3)_H$. As a consequence, we have a smaller value
of $\a_c$ than the prediction of the usual SUSY-GUT's.
We note that the recent experimental values of $\a_c$
\cite{Par}
seem to support the present models.
As for the gaugino masses, we have no prediction without an extra
assumption.

An important ingredient in the present models is the presence of unbroken $R$
symmetry. As a direct consequence of the symmetry,
the dimension-five operators for nucleon decays
\cite{Wei}
are suppressed.
Thus the observation of dimension-five nucleon decays would exclude the present
type of natural unification models.
\footnote{The natural unification model in Ref.\cite{Hot}
allows the dimension-five operator for nucleon decays.}

All of the natural unification models are based on products of two distinct
gauge groups, one of whose coupling constants is in a perturbative regime
and the other in a strong coupling region.
This basic structure of the gauge group might be realized
in some superstring theories.
In fact, a recent development on string theory indicates that nonperturbative
dynamics in string theory may produce some extra gauge symmetries
beside the perturbative one
\cite{Sch},
though compactifications to four-dimensional
spacetime is not yet thoroughly understood.
We hope that an extensive study on the superstring compactification
to realistic models
reveals the structure of unification in elementary particle physics.

\begin{table}[p]
\begin{tabular}{|c||c|c||c|c|}
 \hline
 & \multicolumn{2}{|c||}{First model} & \multicolumn{2}{|c|}{Second model} \\
 \cline{2-5}
 & \makebox[3em]{U$(1)_R$} & \makebox[3em]{U$(1)_\chi$} &
 \makebox[3em]{U$(1)_R$} & \makebox[3em]{U$(1)_\chi$} \\
 \hline
 $Q^i_\a, \ {\bar Q}^\a_i$ & 0 & 1 & 0 & 1 \\
 \hline
 $Q^6_\a, \ {\bar Q}^\a_6$ & 0 & 1 & 2 & $-1$ \\
 \hline
 $X^\a_\b, \ \f$ & 2 & $-2$ & 2 & $-2$ \\
 \hline
 $H_i, \ {\bar H}^i$ & 2 & $-2$ & 0 & 0 \\
 \hline
\end{tabular}\vspace{1em}

\caption{Charge assignments in the two models.}
\label{tbl: charges}
\end{table}

\end{document}